\documentclass[twocolumn,amsmath,amssymb,showpacs,aps,prl,superscriptaddress]{revtex4}
\usepackage{graphicx}% Include figure files
\usepackage{dcolumn}% Align table columns on decimal point
\usepackage{bm}% bold math
\usepackage{epsfig}
\usepackage{subfigure}
\begin{document}

%\preprint{APS/123-QED}

\title{Genetic Programming for Multi-Timescale Modeling}
\author{Kumara Sastry,$^1$ D. D. Johnson,$^1$ David E. Goldberg,$^2$ Pascal Bellon$^1$\\
$^1$Department of Materials Science \& Engineering, and the Fredrick Seitz Materials Research Laboratory.\\ 
$^2$Department of General Engineering\\University of Illinois Urbana-Champaign, Urbana IL.}
%\affiliation{$^1$Department of Material Science \& Engineering, and the Fredrick Seitz Materials Research Laboratory. $^2$Department of General Engineering\\University of Illinois at Urbana Champaign, Urbana IL.}
%\author{D. D. Johnson}%
%\affiliation{Department of Material Science \& Engineering, and the Fredrick Seitz Materials Research Laboratory}
%\author{David E. Goldberg}
%\affiliation{Department of General Engineering\\University of Illinois at Urbana Champaign, Urbana IL.}
%\author{Pascal Bellon}
%\affiliation{Department of Material Science \& Engineering, and the Fredrick Seitz Materials Research Laboratory}

\date{\today}

\begin{abstract}
  A bottleneck for multi-timescale dynamics is the computation of the
  potential energy surface (PES). We explore the use of genetic
  programming (GP) to symbolically regress a mapping of the
  saddle-point barriers from only a few calculated points via
  molecular dynamics, thereby avoiding explicit calculation of all the
  barriers. The GP-regressed barrier function enables use of kinetic
  Monte Carlo (KMC) to simulate real-time kinetics (seconds to hours)
  using realistic interactions. To illustrate, we apply a GP
  regression to vacancy-assisted migration on a surface of a binary
  alloy and predict the diffusion barriers within 0.1--1\% error using
  3\% (or less) of the barriers, and discuss the significant reduction
  in CPU time.
\end{abstract}

\maketitle
\begin{figure}
\center
\epsfig{file=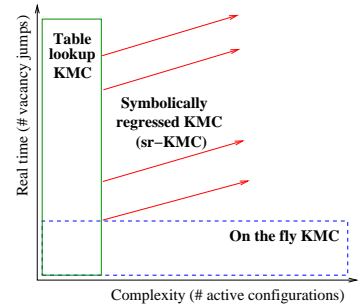,width=1.8in}% Here is how to import EPS art
\caption{\label{fig:KMC} Schematic illustration of simulation
capabilities and bottlenecks of on-the-fly KMC, table look-up KMC, and
symbolically-regressed KMC (sr-KMC).}
\end{figure}

%\section{\label{sec:intro}Introduction}
Molecular dynamics (MD) is extensively used for kinetic modeling of
materials. Yet MD methods are limited to nanoseconds of real time, and
hence fail to model directly many processes. Recently several
approaches were proposed for multiscaling
\cite{Voter:97,Voter:98,Voter:02,Jacobsen:98,Rubia:98,Henkelman:99,Sorensen:00,Cai:02,Steiner:98,Barkema:96,Barkema:00}.
Methods such as temperature-accelerated dynamics (TAD) \cite{Voter:98}
provide significant acceleration of MD but they still fall 3--6 orders
of magnitude short of real processing times. These methods assume that
transition-state theory applies, and concentrate only on infrequent
events.  An alternative approach to bridge timescales
\cite{Jacobsen:98} uses kinetic Monte Carlo (KMC) \cite{Binder:1986}
combined with MD by constructing an {\em a priori} list of events
(i.e., ``look-up table''). The table look-up KMC yields several
orders of magnitude increase in {\em simulated} time over MD (e.g., see 
\footnote{For Cu-Co $\nu_o$ $\approx$ $27\times 10^{12}$~Hz
\cite{Boisvert:97}, giving per time step of KMC relative to MD
(assuming an MD time-step of $10^{-15}$~s) $10^9$ increase in {\em
simulated} time over MD at 300~K, $10^4$ at 650~K, and $10^{2.3}$ at
1000~K}). The table of events is commonly comprised of atomic jumps,
but collective motions (or off-lattice jumps), e.g., see
\cite{Sorensen:00}, may need to be added but may not be known {\em a
priori}. Additionally, tabulating barrier energies from a list of
events is a serious limitation. For example, multicomponent alloys
have an impossibly large set of barriers, due to configurational
dependence, making their tabulation impractical, especially from
first-principles. An alternative approach is calculating energies
``on-the-fly'' \cite{Henkelman:99,Bocquet:2002}, but it too has
serious time limitation (see Fig.~\ref{fig:KMC}). Recent developments
and limitations of KMC methods are given, e.g., in
\cite{Bocquet:2002}.\par

{\em Symbolically-Regressed Table KMC (sr-KMC)}: To avoid the need or
expense of explicit calculation of all activation barriers---frequent
or infrequent---and thereby facilitate an effective hybridization of
MD and KMC for multiscale dynamics modeling, we utilize genetic
programming (GP)---a genetic algorithm that evolves computer
programs---to symbolically regress the PES (in our case saddle-point
barriers only) from a limited set of calculated points on the PES.  An
accurate GP-regressed PES extends the KMC paradigm, as suggested in
Fig.~\ref{fig:KMC}, to machine learn the ``look-up table'' and get an
in-line barrier function for increasing number of active
configurations (or complexity) and providing simulation over
experimentally relevant time frames, which may not be possible from
standard table look-up or on-the-fly KMC.  Interfacing GP with TAD-MD
and/or pattern-recognition methods will further extend its
applicability, e.g., by finding system-specific mechanisms. Of course,
sr-KMC benefits from any advances in KMC methods. In addition,
GP-based symbolic regression holds promise in other multiscaling
areas, e.g., regressing constitutive rules and chemical reaction
pathways, which we are studying.  Also, as we exemplify, standard
basis-set regression are generally not competitive to GP for fixed
accuracy due to the difficulty in choosing appropriate basis
functions.

{\par} To demonstrate, we discuss GP and its application to a
non-trivial case of vacancy-assisted migration on (100) surface of
phase-separating Cu$_x$Co$_{1-x}$.  Although there are millions of
configurations, only the environmental atoms locally around vacancy
and migrating atom significantly influence the barrier energies. We
refer to these as the {\em active} configurations. The results show
that GP predicts barriers within 0.1--1\% error using calculated
barriers of less than 3\% of the total {\em active}
configurations. For alloys, this technique can be combined with a
local cluster expansion technique \cite{Vanderven:01} that reduces the
explicit barrier calculations to $\sim$0.3\% of the {\em active}
configurations.  Our initial results hold promise to enable the use of
KMC (even with realistic potentials) for increased problem complexity
with a scale-up of simulation time.\par

\begin{figure}
\center
\epsfig{file=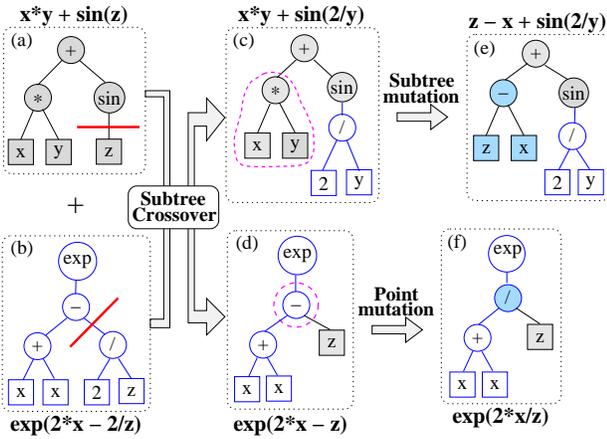,width=3.2in}% Here is how to import EPS art
\caption{\label{fig:GPalgo} Illustration of tree representation,
  subtree crossover, subtree mutation, and point mutation used in
  GP.}
\end{figure}

%\section{\label{sec:GP}Genetic Programming}
{\em Genetic programming} \cite{Koza:92} is a genetic algorithm that
evolves computer programs. The program is represented by a tree
consisting of functions in the internal nodes and terminals in the
leaf nodes (Fig.\ \ref{fig:GPalgo}a). Here we use the function set
${\mathcal{F}} = \{+,-,*,/,\verb+^+,\exp,\sin\}$ and the terminal set
${\mathcal{T}} = \{\vec{x},{\mathcal{R}}\}$, where $\vec{x}$ is a
vector representing the {\em active} alloy configuration, and
${\mathcal{R}}$ is an ephemeral random constant \cite{Koza:92}. Since
we use GP for predicting the barriers, a tree represents a PES-prediction
function that takes a configuration and ephemeral constants as inputs
and returns the barrier for that configuration as
output.\par

A tree's quality is given  by its fitness $f$. For this,
we calculate the barriers $\{\Delta E_{\mathrm{calc}}\left(\vec{x}_1\right), 
%\Delta E_{\mathrm{calc}}\left(\vec{x}_2\right), 
\cdots, \Delta E_{\mathrm{calc}}\left(\vec{x}_M\right)\}$ for $M$ random 
configurations $\{\vec{x}_1,\vec{x}_2,\cdots,\vec{x}_M\}$.  These
configurations are used as inputs to the tree and the barriers 
$\{\Delta E_{\mathrm{pred}}\left(\vec{x}_1\right), 
%\Delta E_{\mathrm{pred}}\left(\vec{x}_2\right), 
\cdots, \Delta E_{\mathrm{pred}}\left(\vec{x}_M\right)\}$, are predicted.
The fitness is then computed as a weighted average of the absolute error 
between the predicted and calculated barriers:
\begin{equation}
\label{eqn:fitness}f = {1 \over M}\sum_{i=1}^{M}w_i\left|\Delta E_{{\mathrm{pred}}}\left(\vec{x}_i\right) -
  \Delta E_{{\mathrm{calc}}}\left(\vec{x}_i\right)\right|
\end{equation}
with $w_i = \left| \Delta E_{\mathrm{calc}}\right|^{-1}$, which gives
preference to accurately predicting lower energy (most significant)
events.\par

Unlike traditional search methods, GP uses a population of candidate
solutions (PES prediction functions) that are initially created using
the {\em ramped half-and-half} method \cite{Koza:92}. Once the population is
initialized and evaluated, the following genetic operators are
repeatedly applied till one or more convergence criteria are
satisfied:\par

\noindent
{\bf Selection:} allocates more copies to solutions with better
  fitness values. We use an {\em $s$-wise tournament selection}
  \cite{Goldberg:89b}, where $s$ candidate solutions are randomly
  chosen and pitted against each other in a tournament. A solution
  with the best fitness wins.\par

\noindent
{\bf Recombination} combines bits and pieces of two solutions to
  create new, hopefully better, solutions.  We use {\em subtree
  crossover} \cite{Koza:92}, where a crossover point for each solution
  is randomly chosen and subtrees below the point are swapped to
  create two new solutions, see Fig.\ \ref{fig:GPalgo}.\par

\noindent
{\bf Mutation} locally but randomly modifies a
  solution. We use two mutation techniques (see Fig.\ 
  \ref{fig:GPalgo}): {\em Subtree mutation}, where a subtree is randomly
  replaced with another randomly created subtree, and {\em point
  mutation} where a node is randomly modified.\par

\begin{figure}[t]
\center
\epsfig{file=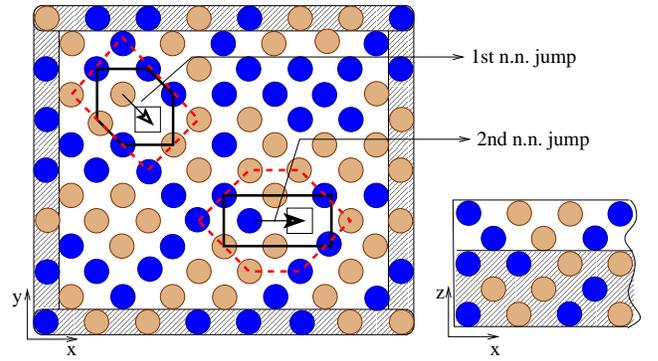,width=3.3in}% Here is how to import EPS art
\caption{\label{fig:configs} Sketch of simulation cell for vacancy-assisted migration on
  (100)-surface of an fcc binary alloy. Atoms in all but the bottom
  layers and the boundary can fully relax. The solid (dashed) lines
  around the migrating atom and vacancy represent 1$^{\mathrm{st}}$
  (2$^{\mathrm{nd}}$) n.n environmental atoms.}
\end{figure}

{{\em Case Study:\/}} We consider the prediction of diffusion barriers
for vacancy-assisted migration on (100) surface of phase-separating
Cu$_x$Co$_{1-x}$. The system consists of five layers with 100 to 625
atoms in each layer (see Fig.\ \ref{fig:configs}). The bottom three
layers are held fixed to their bulk bond distances, while the top
layers are either held fixed (as a test) or fully relaxed via MD. We
consider only first and second nearest-neighbor (n.n.) jumps, along
with 1$^{\mathrm{st}}$ (as a test) and 2$^{\mathrm{nd}}$ n.n.
environmental atoms in the active configuration, as shown in Fig.\
\ref{fig:configs}. This system already exhibits large complexity and
is still small enough so that table look-up and GP-regressed KMC can
be implemented and directly compared. Table\ \ref{tab:table1} gives
the number of active configurations when 1$^{\mathrm{st}}$ and
2$^{\mathrm{nd}}$ n.n. environments are considered for a binary
alloy.\par

\begin{table}[b]
\caption{\label{tab:table1}Number of active configurations
  for 1$^{\mathrm{st}}$ and 2$^{\mathrm{nd}}$ n.n. jumps, and for
  1$^{\mathrm{st}}$ and 2$^{\mathrm{nd}}$ n.n. active atoms}
\begin{ruledtabular}
\begin{tabular}{ccc}
&1$^{\mathrm{st}}$ n.n. jumps&2$^{\mathrm{nd}}$ n.n. jumps\\\hline
1$^{\mathrm{st}}$ n.n. active configurations & 128 & 128\\
2$^{\mathrm{nd}}$ n.n. active configurations & 2048 & 8192\\
Total configurations & $\gg 2^{100}$ & $\gg 2^{100}$
\end{tabular}
\end{ruledtabular}
\end{table}
\begin{figure}[tbh]
\center
\epsfig{file=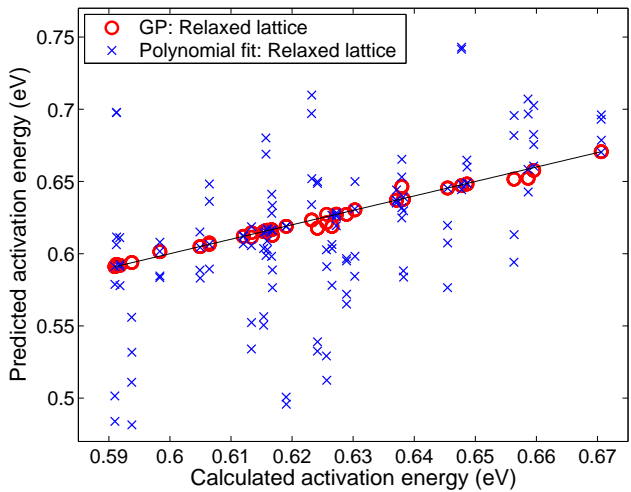,width=3.3in}
\caption{\label{fig:morse1nnR} Activation energies (in eV) predicted
  by regression. GP (circles) and a quadratic polynomial (crosses) are
  compared to the calculated (Morse) barriers for 1$^{\mathrm{st}}$
  n.n. jumps on (100)-surface of Cu$_{0.5}$Co$_{0.5}$ for relaxed
  lattices. As a simple test, only first n.n. environments are
  considered in the active configuration. The line is a guide for the
  eye.}
\end{figure}

We model the atomic interactions with a simple Morse potential
\cite{Girifalco:59} and a tight-binding potential
with second-moment approximation (TB-SMA) \cite{Levanov:2000}. To
validate interactions, we model vacancy-assisted migration on
(100)-surface of Cu and consider only first n.n. jumps. The predicted
barrier for n.n. vacancy jumps with fully relaxed lattice in Cu is
0.39~eV for Morse and 0.45~eV for TB-SMA, agreeing with 0.42$\pm$0.08
(0.47$\pm$0.05) from {\em ab initio} (EAM) \cite{Boisvert:97}
calculations.\par

We now consider the barrier regression via GP for vacancy-assisted
migration on (100)-surface of Cu$_x$Co$_{1-x}$. The input to the
barrier regression (i.e., prediction) function, $\vec{x} =
\left\{x_j\right\}$ is a binary-encoded vector sequence, where $x_j =
0~(1)$ represents a Cu (Co) atom. For simplicity, we begin by
considering only seven 1$^{\mathrm{st}}$ n.n. environmental atoms
yielding 128 {\em active} configurations. About 20, i.e., 16\%,
different active configurations are randomly chosen and their barriers
are computed using the conjugate-gradient method and are used in the
GP fitness function, see Eq.~\ref{eqn:fitness}. The barriers
predicted by GP for the relaxed configurations are compared to the
exact values in Fig.\ \ref{fig:morse1nnR}. We note that the prediction
error for rigid lattice case (0.4$\pm$0.04\%) is significantly less
than that for relaxed lattice case (2.8$\pm$0.08\%). Due to the
weighting used in the fitness function, GP predicts barriers for most
significant events more accurately than for less-significant
(higher-energy) events.\par

Figure~\ref{fig:morse1nnR} also compares the barriers predicted by GP
to those predicted by a least-squares fit quadratic polynomial,
showing clearly its inadequacy for alloys.  Furthermore, while GP
requires only 16\%, the quadratic (cubic) polynomial fit needs 27\%
(78\%) of the barriers.  In limited cases, such as dilute
Fe$_{1-x}$Cu$_x$, the barriers can be predicted via a simple
polynomial fit \cite{Bouar:2002}.\par

\begin{figure}
\center
\epsfig{file=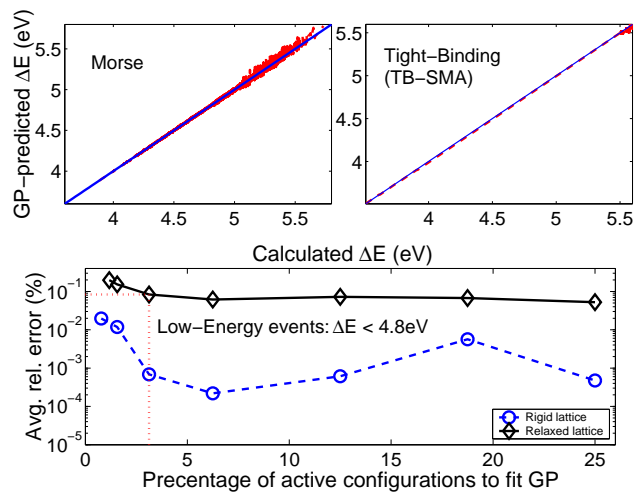,width=3.3in}
\caption{\label{fig:morseTB2nnRe} (Upper) Calculated vs. GP-predicted
  barriers (in eV) for 2$^{\mathrm{nd}}$ n.n. jumps on relaxed
  (100)-surface of Cu$_{0.5}$Co$_{0.5}$ active configurations up to
  2$^{\mathrm{nd}}$ n.n. for Morse and TB-SMA.
  (Lower) GP predicts the barriers with 0.1\% (1\%) error for
  most- (less-) significant events with $\Delta E <$ 4.8~eV
  ($\Delta E >$ 4.8~eV) from only 3\% of active configurations.}
\end{figure}

To test the scalability of GP with {\em active} configuration size, we
consider the 2$^{\mathrm{nd}}$ n.n. jumps and 1$^{\mathrm{st}}$ and
2$^{\mathrm{nd}}$ n.n. environmental atoms in the {\em active}
configuration. As shown in Table~\ref{tab:table1}, there are a total
of 8192 configurations. The energies predicted by GP are compared with
direct calculations in Fig.\ \ref{fig:morseTB2nnRe}, along with the error.
The GP predicts the barriers for most significant events with less
than 0.1\% error by fitting to energies from only 3\% (i.e., 256/8192)
of the active configurations (with error defined in \footnote{The
average relative error for $N_{\mathrm{cfgs}}^{\prime}$ configurations
within the desired energy range is given by
\[
\bar{\varepsilon}_{{\mathrm{rel}}} = {100 \over
  N_{\mathrm{cfgs}}^{\prime}}\sum_{i = 1}^{N_{\mathrm{cfgs}}^{\prime}} \left| {\Delta E_{\mathrm{pred}}\left(\vec{x}_i\right)
  - \Delta E_{\mathrm{calc}}\left(\vec{x}_i\right) \over \Delta E_{\mathrm{calc}}\left(\vec{x}_i\right)}\right|,
\]}). In comparison, a cubic polynomial fit requires energies for $\sim$6\% of the
configurations, predicting the barriers with 2.5\% error for the most
significant events.

The results shown in Figs.\ \ref{fig:morse1nnR} and
\ref{fig:morseTB2nnRe} clearly demonstrate the effectiveness of GP in
predicting the potential energy surface, with high accuracy and little
information.  As expected, since the regression and barrier
calculation are nearly independent, the GP performance does not depend
on the potentials used, e.g., Fig.~\ref{fig:morseTB2nnRe} shows
results for both Morse, and non-additive and non-linear tight-binding
potentials. The regression only requires a database of barriers and
has no knowledge (nor the need) of the underlying potential used. We
also find that the GP performance is independent of the configuration
set used in calculating the fitness function, the order in which they
are used, and the labeling scheme used to convert the configuration
into a vector of inputs. Differences in activation-energy scale on the
PES prediction via GP is also negligible. That is, even though the
barriers for the $1^{\mathrm{st}}$ and $2^{\mathrm{nd}}$ n.n. jumps
differ by an order of magnitude, GP predicts the barriers with similar
accuracy.  Moreover, for more complex, cooperative effects, such as
island diffusion via surface dislocations \cite{Hamilton:95}, sr-KMC
could be interfaced with pattern-recognition methods \footnote{Talat
Rahman, private communication} (see \footnote{For long-range fields
(e.g., elastic fields from coherent interfaces, such as multilayers or
precipitates), a description based solely on local configurations may
have to be extended, say, with phase field methods.} for long-range
fields).\par

Time enhancements by coupling a GP-regressed barriers with KMC (or
sr-KMC) are simple to estimate. For our example, with $\sim$33 times
fewer calculated barriers GP symbolically regresses an in-line barrier
function---rather than the complete look-up table---and thus, sr-KMC
provides a direct CPU savings of $\sim$100 over table look-up
methods. Additionally, each time step of sr-KMC requires only
$10^{-3}$ CPU-seconds for an in-line function evaluation, as opposed
to on-the-fly KMC which require seconds (empirical potentials) to
hours (quantum methods), thus providing a gain of 10$^4$--10$^7$
CPU-seconds. For our example, one relaxed barrier calculation takes
$\sim$10 secs ($\sim$1800 secs) for Morse (TB-SMA). One important
question, especially for bulk diffusion, is how the gain from sr-KMC
scales with system complexity. While we cannot fully answer this
question yet, in the present study it is remarkable and promising that
the fraction of explicit barrier calculations required by sr-KMC
decreases as the number of active configurations increases.\par

To summarize, potential energy surface prediction using symbolic
regression via genetic programming (GP) holds promise as an efficient
tool for multi-scaling in dynamics. The GP based KMC approach avoids
the need or expense of calculating the entire potential-energy
surface, is highly accurate, and leads to significant scale-up in
simulation time and a reduction in CPU time.  We have shown on a
non-trivial example of vacancy-assisted migration on a surface of
Cu$_x$Co$_{1-x}$ that GP predicts all barriers with 0.1--1\% error
from calculations for only 3\% of active configurations, allowing
seconds of simulation time. For alloy problems, the number of direct
barrier calculations can further be reduced by over an order of
magnitude by hybridizing GP with cluster expansion methods
\cite{Vanderven:01}. We emphasize that the GP is non-trivially
regressing a function and its coefficients that approximates the
potential-energy surface, and its efficacy over standard basis-set
regression is clear. Moreover, GP approach is not problem specific and
requires little modification, if any (say by choice of operators and
functions), to address increasingly complex cases, and as suggested,
is potentially useful in other multiscaling areas.

\begin{acknowledgments}
This work was supported by the NSF at Materials Computation Center
(ITR grant DMR-99-76550), CPSD (ITR grant DMR-0121695), AFOSR (grant
F49620-00-0163), Computational Science \& Engineering fellowship and
the Dept. of Energy through the Fredrick Seitz MRL (grant
DEFG02-91ER45439) at UIUC. We gratefully acknowledge discussions with
R.S. Averback, Y. Ashkenazy, and J.-M. Roussel.
\end{acknowledgments}
\bibliography{GPpredictsPES}% Produces the bibliography via BibTeX.

\end{document}